\documentclass[conference]{IEEEtran}
\IEEEoverridecommandlockouts
\usepackage{subcaption}
\usepackage{cite}
\usepackage{amsmath,amssymb,amsfonts}
\usepackage{algorithmic}
\usepackage{graphicx}
\usepackage{textcomp}
\usepackage{xcolor}
\usepackage{url}
\usepackage[stable]{footmisc}
\def\BibTeX{{\rm B\kern-.05em{\sc i\kern-.025em b}\kern-.08em
    T\kern-.1667em\lower.7ex\hbox{E}\kern-.125emX}}
\begin{document}

\title{Towards Performance Portable Programming for Distributed Heterogeneous Systems}

\author{\IEEEauthorblockN{Polykarpos Thomadakis}
\IEEEauthorblockA{\textit{Department of Computer Science} \\
\textit{Old Dominion University}\\
Norfolk, Virginia \\
pthom001@odu.edu}
\and
\IEEEauthorblockN{Nikos Chrisochoides}
\IEEEauthorblockA{\textit{Department of Computer Science} \\
\textit{Old Dominion University}\\
Norfolk, Virginia \\
nikos@cs.odu.edu}
}

\maketitle

\begin{abstract}
Hardware heterogeneity is here to stay for high-performance computing. Large-scale systems are currently equipped with multiple GPU accelerators per compute node and are expected to incorporate more specialized hardware in the future. This shift in the computing ecosystem offers many opportunities for performance improvement; however, it also increases the complexity of programming for such architectures. This work introduces a runtime framework that enables effortless programming for heterogeneous systems while efficiently utilizing hardware resources. The framework is integrated within a distributed and scalable runtime system to facilitate performance portability across heterogeneous nodes. Along with the design, this paper describes the implementation and optimizations performed, achieving up to 300\% improvement in a shared memory benchmark and up to 10 times in distributed device communication. Preliminary results indicate that our software incurs low overhead and achieves 40\% improvement in a distributed Jacobi proxy application while hiding the idiosyncrasies of the hardware. 
\end{abstract}


\section{Introduction}
The recent slowdown in Moore's Law is leading to large-scale disruptions in the computing ecosystem. Users and vendors are transitioning from utilizing computing nodes of relatively homogeneous CPU architectures to systems led by multiple GPU devices per node. This trend is expected to continue in the foreseeable future, incorporating many more types of heterogeneous devices, including FPGAs, System-on-Chips (SoCs), and specialized hardware for artificial intelligence\cite{osti_1822199}. The new computing ecosystem sets the basis to significantly improve performance, energy efficiency, reliability and security; thus, high-performance computing (HPC) systems are adapted to optimize their performance on traditional workloads and modern workloads.      

Exploiting extreme heterogeneity requires the development of new techniques and abstractions that handle the increased complexity in productivity, portability and performance. The new techniques should allow users to express their applications' workflow in a uniform way, hiding the idiosyncrasies of the underlying architecture while implicitly handling performance portability by optimizing scheduling, load balancing and data transfers between the heterogeneous devices of a node. Utilizing and orchestrating data movement and work on multiple such nodes, further increases the complexity of developing heterogeneity-aware applications. Thus, the runtime system should also facilitate seamless use of distributed heterogeneous nodes, by providing abstractions for data and workload that are independent of the underlying hardware.

In this paper, we present a novel runtime system that enables seamless, efficient and performance portable development of distributed heterogeneous applications. First, we introduce a heterogeneous tasking framework that aims to optimize the parallel execution of heterogeneous tasks on a single node. The tasking framework provides a programming model that automatically leverages heterogeneous devices. In contrast to other systems, our framework does not require the application to pick a device where a task should run; instead, the application only picks a device type and the framework is responsible to schedule the task to the optimal device. Second, we extend a homogeneous distributed system, namely the Parallel Runtime Environment for Multicomputer Applications (PREMA)~\cite{Thomadakis22M}, to handle heterogeneous devices by integrating it with the heterogeneous tasking framework. 
Along with the design and implementation of the final product, we present optimizations that contributed to achieving high performance. The evaluation results with microbenchmarks and a proxy application show that our system incurs low overhead with scalable performance.

The major contributions of this paper are as follows.
\begin{itemize}
    \item A stand-alone tasking framework offering performance portability over multiple heterogeneous devices.
    \item Integration of
    the aforementioned tasking framework into a distributed runtime to leverage complex distributed heterogeneous computing systems.
    \item A series of performance optimizations that achieve significant improvements.
    \item Performance analysis on a distributed proxy application.
    
\end{itemize}

\section{Background}
\label{sec:background}
The Parallel Runtime Environment for Multicomputer Applications (PREMA)~\cite{Thomadakis22M} is a runtime framework, designed to handle the needs of applications targeting extreme-scale homogeneous computing platforms. It manages the burden of latency-hiding, shared and distributed memory scheduling/load balancing, and provides a global address space that drastically decreases the complexity of developing extreme-scale applications.
PREMA consists of three software layers that provide different features according to the principle of separation of concerns. 

The first layer, the \textbf{ Data Movement and Control Substrate (DMCS)}, offers an asynchronous message-driven execution model where messages are associated with a task/function (referred to as handler in this context) that is invoked implicitly on the receiver upon their arrival, similar to Active Messages~\cite{vonEicken92AM}. Communication and handler invocations are by default asynchronous, and one-sided, requiring no explicit involvement of the receiver. 
The \textbf{Mobile Object Layer (MOL)} extends DMCS with the introduction of the \emph{mobile objects}. A mobile object is a location-independent container that is provided by the runtime system to capture coarse-grained application data. Mobile objects can be targeted for remote handler executions uniformly and independently of their location. 
Adapting to a programming model where the workflow is expressed as interactions (i.e., handler invocations) between mobile objects leads to a more naturally asynchronous design for the application and allows PREMA to better handle latencies while exposing a uniform programming interface that hides the structure of the underlying platform. Mobile objects are also used to provide applications with implicit distributed load balancing through the \textbf{Implicit Load Balancing (ILB)} layer but this feature is not utilized in the context of this work.


In order to provide more flexibility for the runtime system to overlap latencies, an application is encouraged to perform over-decomposition of its data. In over-decomposition, the data domain of an application running on a platform of $P$ processing elements is decomposed into $N$ chunks where $N \gg P$. Decomposing the data domain into many more pieces than the available processing elements gives the runtime system more options on scheduling computational tasks and filling the idle time stemming from data movement and communication operations.   

\section{Related Work}
\label{sec:related_work}
Several systems have been adapted to efficiently utilize GPUs in their workflow, while new ones have emerged in an attempt to create new standards for their use. Systems like Charm++~\cite{Charm++}, HPX~\cite{HPX} and X10~\cite{X10} have introduced new interfaces to provide support for GPUs. However, these systems let the users explicitly handle issues like requesting memory transfers, managing device platforms, task allocations, work queues to optimize performance. In contrast, the proposed work provides a uniform abstraction for heterogeneous tasks and data, and implicitly handles scheduling, load balancing and latency overlapping independently of the target device backend. StarPU~\cite{StarPU}, OmpSs~\cite{ompss} and ParSec~\cite{ParSec} offer different high-level approaches to efficiently utilize distributed heterogeneous systems. However, their programming model is mostly suitable for applications whose workflow follow a regular pattern and can be inferred mostly statically. PREMA, on the other hand, adopts a dynamic, message-driven programming model that is more suitable for irregular applications. 

SYCL and DPC++/oneAPI~\cite{DPC++}, as well as the newest version of OpenMP are recent attempts to  provide performance portable interfaces in modern C++ that can target heterogeneous devices. However, users still need to handle load balancing, scheduling and work queues for multi-device systems and need to combine them with another runtime solution that targets distributed nodes. In fact, such systems could be implemented with our heterogeneous tasking framework as a interoperability backend.

\section{Design and Implementation}
\label{sec:des_impl}

\subsection{Programming Model}
The programming model of the heterogeneous tasking framework builds upon two simple abstractions: the heterogeneous objects (\emph{hetero\_objects}) and heterogeneous tasks (\emph{hetero\_tasks}). A \emph{hetero\_object} uniformly represents a user-defined data object residing on one or more computing devices of a heterogeneous compute node (e.g., CPUs, GPUs, FPGAs). Applications treat such objects as opaque containers for data without being aware of their physical location. A \emph{hetero\_task} encapsulates a non-preemptive computing kernel that runs to completion and implements a medium-grained parallel computation. Like hetero\_objects, hetero\_tasks are defined and handled by the application in a uniform way, independent of the device they will execute on.

\subsubsection{Heterogeneous Objects} 
Handling copies of the same data on different heterogeneous devices can lead to error-prone, and difficult to understand and maintain application code. In general, applications need to handle data transfers between them, use the correct pointer for the respective device and also keep track of their coherence. A hetero\_object is an abstraction that automatically handles such concerns, maintaining the different copies of the same data in a single reference.  The underlying system handles hetero\_objects to guarantee that the most recent version of the data will be available at the target device at the time that they will be needed. For example,
accessing an object originally located on the host from a  GPU would automatically trigger the transfer of the underlying data from the host to the respective device. In the same manner, accessing the same object from a different device would trigger a transfer from the GPU to that device. The runtime system  guarantees data coherence among computing devices, keeping track of up-to-date or stale copies and handling them appropriately.

The actual data captured by a hetero\_object should mainly be accessed and modified through hetero\_tasks for optimal performance. However, the application can also explicitly request access to the underlying data on the host after specifying the type of access requested, in order to maintain coherence. This method will trigger (if needed) an asynchronous transfer from the device with the most recent version of the data and immediately return a future. The future can then be used to query the status of the transfer and provide access to the raw data. In this state, the data of the hetero\_object are guaranteed to remain valid on the host side, preventing tasks that would alter them from executing, until the user explicitly releases their control back to the runtime system. Since the application has no direct access to the actual memory that is allocated in different devices, our framework monitors the memory usage of each device. When a device memory is close to be depleted, the runtime system will start offloading some of the users data to the host or other devices automatically. We currently use a Least Recently Used (LRU) policy to decide which hetero\_object should be offloaded to free space in a device. In order to avoid memory over-consumption an application can explicitly request for a hetero\_object to be completely removed from all devices, however, hetero\_objects will also be automatically freed when going out of scope. 

\subsubsection{Heterogeneous Tasks} 
Heterogeneous tasks (hetero\_tasks) are opaque structures that are used to consolidate the parameters that characterize a computational task. Through a hetero\_task applications define the kernel to execute, input/output data arguments, processing elements requested (e.g., threads in a CPU), task dependencies and type of target device. Moreover, applications can request the allocation of a temporary shared memory region available only for the duration of the kernel which maps to the concept of local/shared memory found in other GPU programming APIs (CUDA, OpenCL).

Heterogeneous tasks are independent of the underlying target hardware, allowing a uniform expression of the application workflow whether they target CPUs, GPUs, or other types of devices. The computational kernel that they represent is defined in a dialect very similar to an OpenCL kernel which is translated appropriately for the target device. The input/output data arguments of a task are defined as the hetero\_objects that it needs to access, along with the type of access that is required for each of them (read, write). This information is used by the runtime to issue the appropriate data transfers, maintain coherence and also infer task dependencies.
Submitting a tasks for execution does not immediately execute the respective kernel; instead, the runtime system defers the task execution for a later point and immediately returns control to the user. The application, however, can query whether the kernel execution has been completed or wait for its
completion. Moreover, The heterogeneous tasking framework provides methods for handling task dependencies both explicitly and implicitly.

Applications can \textbf{explicitly define a task dependency graph} using the \emph{add\_dependency()} method of hetero\_tasks. Once their dependencies have been set, the application can submit the respective tasks for execution all at once. 
This approach allows the runtime system to improve performance while removing much of the burden of guaranteeing correctness from the application. 
To further reduce the effort required to guarantee correctness, the framework also supports \textbf{implicit task dependency detection} based on the arguments accessed by each hetero\_task. Assuming that the application submits tasks in a correct sequential order, conflicting tasks are guaranteed to execute in the correct order while independent tasks will automatically explore maximum parallelism and avoid race conditions.


\subsection{Execution Model}
Heterogeneous tasks are executed in an asynchronous manner from the tasking framework. When a task is submitted for execution, it is appended to a list of task execution requests. The control is immediately returned back to the application which can continue to submit more tasks or execute other work. A separate component of the runtime (optionally running in a separate thread) examines task execution requests and eventually schedules them for execution after performing a number of steps that guarantee correctness.

The first step towards executing a task is to infer its data dependencies with other tasks based on its data arguments. To achieve this, the runtime maintains a list of the currently submitted or running tasks that target each hetero\_object and automatically sets the dependencies of new tasks that conflict with the ones already submitted or running. Tasks with at least one incomplete dependency are pushed to another queue of blocked tasks, otherwise they are appended directly to the scheduler's active work-pool. Tasks in the blocked queue are periodically evaluated by the scheduler to infer if there dependencies have been fulfilled, in which case they are moved to the scheduler's pool of runnable tasks.

Once all blocked tasks have been examined, the scheduler is ready to schedule the runnable tasks. At this stage, the scheduler decides the order in which the tasks should execute and the device where they should run, based on the user's device type preference (i.e., the scheduler chooses the specific device ID while the user only gives a preference for a device type). The runtime will then reserve device resources and issue the data transfer requests of the input/output hetero\_objects to be accessed on the chosen device. Moreover, it will automatically try to overlap the different operations, if possible, by utilizing the features provided by the target device's API (e.g., CUDA streams or OpenCL command queues). When all outstanding data transfers of a task have completed, the computational kernel will be submitted to the target's work-pool. Submitted tasks are periodically checked for completion by the runtime to update the status of pending dependent tasks.

\subsection{Scheduler}
With the introduction of more heterogeneous compute devices and workloads, it is expected that scheduling and load balancing will only become more complicated. To allow more flexibility for different use cases, the actual implementation of the scheduler is designed to be modular and separate from the rest of the heterogeneous tasking framework. We provide the scheduler as an abstract class that only requires two operations to be implemented. The \emph{push()} operation is responsible to enqueue a new runnable task into the scheduler's work-pool while the \emph{pop()} operation returns the next task that should be executed as well as the device it should run on. The abstract scheduler class allows the development of as simple or complex custom data structures and policies as the user might need. 

\subsection{Implementation}
The heterogeneous tasking framework is implemented in the C++ programming language leveraging from its performance and its object oriented design. It is developed in three software layers in order to allow easy integration with new device types, programming APIs and scheduling policies. 

The \textbf{Device API} is the bottom layer that abstracts the different operations provided by a heterogeneous device vendor. It consists of abstract C++ classes that expose virtual methods for operations required to (a)synchronously issue tasks and manage data in such devices,
as well as methods to query the status of an asynchronous operation and the hardware specifications of a device. Currently we provide native support with CUDA and OpenCL for GPUs. Applications have the option to allocate a dedicated thread that handles the requests for each device or have a single thread that manages all devices.


The next layer is the \textbf{Core Runtime} layer which encapsulates the underlying implementation of the hetero\_objects and hetero\_tasks, monitors the coherence of the different copies of the data and detects and enforces task dependencies. It makes use of the Device API to coordinate data transfers, guide the correct execution of tasks and signal the completion of different operations.
This is the layer that acts as the "glue" between the application preferences, the scheduler and load balancing policies and the Device API.

At the top stands the \textbf{Application Layer} which consists of a thin API that exposes the capabilities of the tasking framework in a high-level interface. 
In the current implementation, kernels are defined in a dialect similar to OpenCL through the use of macros which are then expanded to define a version of the kernel for each available target. 

\section{Heterogeneity within PREMA}
\label{sec:prema_integration}
Adding heterogeneity in PREMA is an important requirement that has to be completed in 
order to handle the load of exascale-era machines. Applications should be given the ability to use and transfer device memory in the context of remote handler executions without much hassle. A step towards this direction is to allow PREMA to send and receive buffers located in a GPU device either explicitly (currently CUDA only) or through the abstractions of the heterogeneous tasking framework we have introduced. The explicit approach allows users to utilize GPUs without restricting them to utilize our framework, in order to facilitate interoperability with legacy CUDA codes. It also provides a barebone approach to integrate heterogeneity on top of the distributed system that will also act as the base case for our performance evaluation since it introduces the least possible overhead.

\subsubsection{Explicitly Handling Devices}
In the explicit approach, the application can directly call the different GPU operations of the CUDA API to allocate/free memory, initiate transfers and execute kernels. In this approach, PREMA provides an API to invoke remote handlers that include a GPU buffer as an argument. This API includes  the ID of the remote process, the buffer to be transferred, its size, the IDs of the source and target devices and the handler (host function) that should be invoked at the receiver. PREMA will handle the transfer of the buffer between the remote GPUs and invoke the handler when the buffer has been copied to the target GPU. The handler is then free to invoke any GPU-related operation that targets this buffer safely. However, the application needs to make sure that the handler does not return before the completion of the kernel, since any buffers transferred through a handler will be removed at its return, including the GPU buffer. Waiting for all the device operations to complete (e.g., through \emph{cudaDeviceSynchronize()}), is enough to guarantee correctness, however, this approach will harm PREMA's time-slicing abilities, preventing it from switching to other tasks while GPU operations are in progress.  Thus, the user should follow a more complicated approach, querying the status of the operations without blocking (e.g., through \emph{cudaEvents}) and periodically yielding control of the thread for PREMA to run background jobs.

\subsubsection{Utilizing the Heterogeneous Tasking Framework}
To facilitate a higher-level interaction of PREMA with heterogeneous devices, we introduced a set of extensions that allow it to directly utilize the abstractions provided by the heterogeneous tasking  framework. 
As a comparison with the explicit remote handler invocation API, the user now only needs to provide the handler to be executed, the target process ID and the hetero\_object to be passed as argument. 
Since the hetero\_objects handle the location of the underlying data the user does not need to specify their location, moreover, the framework automatically decides the device to store the received buffer on the target process. Once a hetero\_object of a remote method invocation has been transferred, the designated handler is invoked on the target and the application can invoke tasks that utilize it on any type of device that is available. Waiting for such tasks to terminate through the appropriate API does not impact the performance of PREMA since the respective functionality has been  adapted to allow implicit time-slicing.

Another desirable requirement that is provided through the use of hetero\_objects on top of PREMA is the ability to ``put'' and ``get'' data between potentially distributed devices. A new extension allows users to create global pointers for hetero\_objects, i.e., unique identifiers referenceable from all processes in the distributed system. 
When an application needs to store/retrieve data to/from a remote hetero\_object it just needs to provide its global pointer and the location (hetero\_object or pointer) of the data to be read/written, along with a callback that is triggered on the target, signaling the completion of the operation.   

\subsubsection{Implementation}
\label{sec:hetero_prema_impl}
Depending on the capabilities of the underlying communication library and the target device hardware the actual implementation of the memory transfers differs to leverage heterogeneity-aware communication substrates.
Implementing the memory transfers when the application explicitly handles device memory and  host staging is used includes the following steps:
\begin{enumerate}
    \item The sender prepares a header encapsulating  the handler invocations information, e.g. data, target device ID, etc.
    \item The sender allocates host memory, copies the device data there and sends the two buffers
    \item The receiver detects that a new message is arriving and allocates a new memory to store the header.
    \item The receiver allocates host memory for the data and posts a receive operation.
    \item The receiver allocates memory on the specified device, copies the buffer there and invokes the handler.
\end{enumerate}
In contrast, when PREMA's hetero\_objects are utilized the process is similar but also includes packing additional information stored in the hetero\_object for step 1), accessing the host-side data through the framework in 2) which includes operations to guarantee that the buffer is not written before the send has completed, unpacking the hetero\_object additional information and requesting the framework to allocate device memory based on the data buffer received on the host in 4).
The process to implement the put/get data is almost the same but it includes a step to lookup the location of the hetero\_object on both the sender and the receiver and also, instead of allocating new memory on the receiver device the received buffer is directly copied to the target hetero\_object through the tasking framework interface.

If the communication library/hardware and compute devices support direct transfers between distributed devices (currently only tested for CUDA-OpenMPI) one can skip the intermediate host staging step. Thus, an implementation for the case where the user explicitly handles device memory is the following:
\begin{enumerate}
    \item The sender prepares a header that encapsulates information regarding the handler invocation.
    \item The sender first sends the message header and then the device buffer directly without any copies. 
    \item The receiver detects that a new message is arriving and allocates a new memory to store the header.
    \item The receiver allocates memory on the specified device, receives the second message directly in the device memory and invokes the designated function (handler).
\end{enumerate}
In contrast, when PREMA's hetero\_objects are utilized the process is similar but also includes packing additional information stored in the hetero\_object for step 1), accessing the device-side data through the framework in 2) which includes operations to guarantee that the buffer is not written before the send has completed, unpacking the hetero\_object additional information and requesting the framework to allocate device memory and return it in 4). The process to implement the put/get data is almost the same but it includes a step to lookup the location of the hetero\_object on both the sender and the receiver and also, instead of allocating new memory on the receiver device the hetero\_object's buffer is first retrieved through the framework's interface and the buffer is directly received there.

\section{Performance Evaluation }
\label{sec:perfom_opt}
This section investigates optimizations that can improve the performance of different operations provided by the heterogeneity aware PREMA and the tasking framework and presents the performance of the optimized implementation on a proxy application \footnote{Benchmarks available at: \url{https://git.cs.odu.edu/cs_pthom001/heterogeneous-prema}}. We used a small cluster consisting of 16 32-core machines, which have two Intel Xeon Gold 6130 CPUs (2.1 GHz) and four NVIDIA Tesla V100 GPUs. 

\subsection{Heterogeneous Tasking Framework}
We evaluate different performance optimization techniques on this framework for NVIDIA GPUs using a simple, double precision matrix-matrix multiply benchmark, and attempt to optimize the mean throughput achieved out of 100 iterations. In each iteration, the three matrices are allocated in the device, the data are transferred and the compute kernel is executed. Note that the results are not copied back to the host to prevent the CUDA implementation from blocking. We incrementally apply optimizations and compare against a simple CUDA implementation of the benchmark as our baseline, when using a dedicated thread for the device and without it.  As can be seen in Fig.~\ref{fig:dgemm_benchmarks}, the baseline bars in both cases show that the framework adds some overhead on top of the naive CUDA implementation which is signficant especially for the case where a dedicated thread is used. PREMA overcomes some of these overheads with the optimizations presented

\subsubsection{Page-locked Host Memory}
In order to fully utilize the bandwidth capabilities of the respective hardware, the NVIDIA GPUs require that host data that are transferred to the GPU should reside in a page-locked memory region. Moreover, this is the only way that device to host transfers can be asynchronous with respect to the host. To accommodate this, applications need to explicitly (de)allocate memory in a special way, which increases code complexity and induces an overhead much higher compared to regular memory allocations. 

For this purpose, at the initialization step of the framework, we allocate a large chunk of page-locked memory that is later used as a memory-pool for host memory allocations. 
This optimization gives a huge boost that almost doubles the performance of the framework when no dedicated thread is used. On the other hand, when utilizing a dedicated thread the improvement is much higher  (80\%) for large matrices than smaller ones (30\%) (Fig.\ref{fig:dgemm_benchmarks}; TF-Pagelocked). When investigating the performance degradation observed for the dedicated thread implementation with the profiler, we noticed that asynchronous GPU operations (e.g., cudaMemcpyAsync()) would take longer to return when issued from the dedicated thread. Work is in progress to understand and alleviate this issue. 

\subsubsection{Custom Device Memory Pools}
\label{sec::dev_mem_pool}
Allocating and freeing device memory are expensive operations that require synchronization
between the host and the device. Moreover, the two functions might require the completion
of previously issued asynchronous operations before running. The proposed runtime system
uses a custom memory allocator per device to avoid the overheads stemming from constantly
requesting new memory (de)allocations. During the initialization of the runtime system, a
request to allocate most of the available memory of each device (except the host) is issued,
and the returned memory is handled by the custom memory allocator. When memory
needs to be allocated, the custom allocator is used instead of the one provided by the device library.
Implementing this optimization improved performance by up to 10\% in default implementation where no dedicated thread is utilized but provided a significant improvement in the dedicated thread case.  Specifically, for smaller matrices ($\leq$ 256x256) we observe an improvement of about 100\%; for larger matrices the improvement is about 20\% (Fig.\ref{fig:dgemm_benchmarks}; TF-CustomAlloc).


\subsubsection{Enabling Concurrent GPU Operations} 
\begin{figure}[t]
    \centering
    \includegraphics[width=0.8\linewidth]{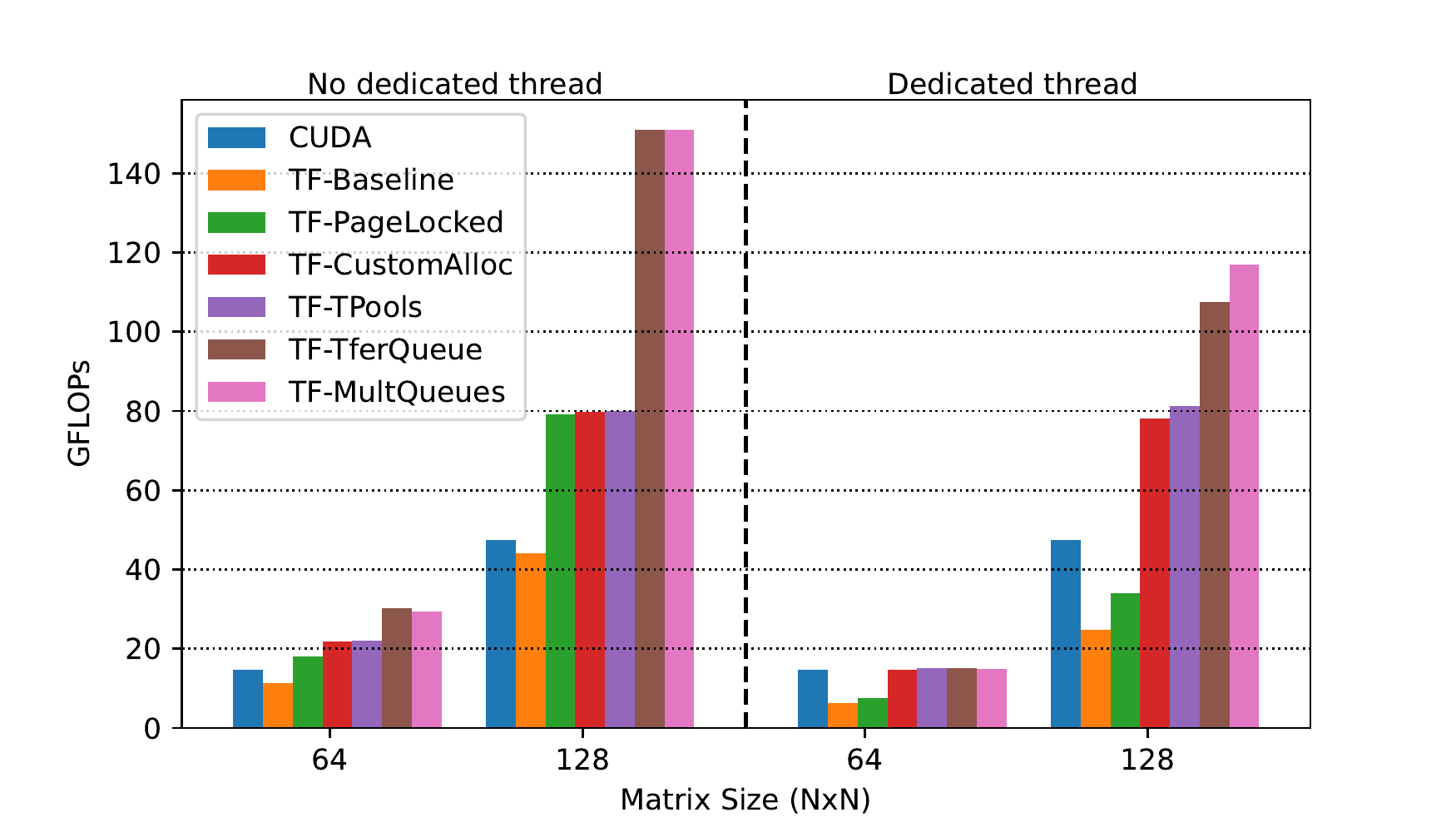}
    \includegraphics[width=0.8\linewidth]{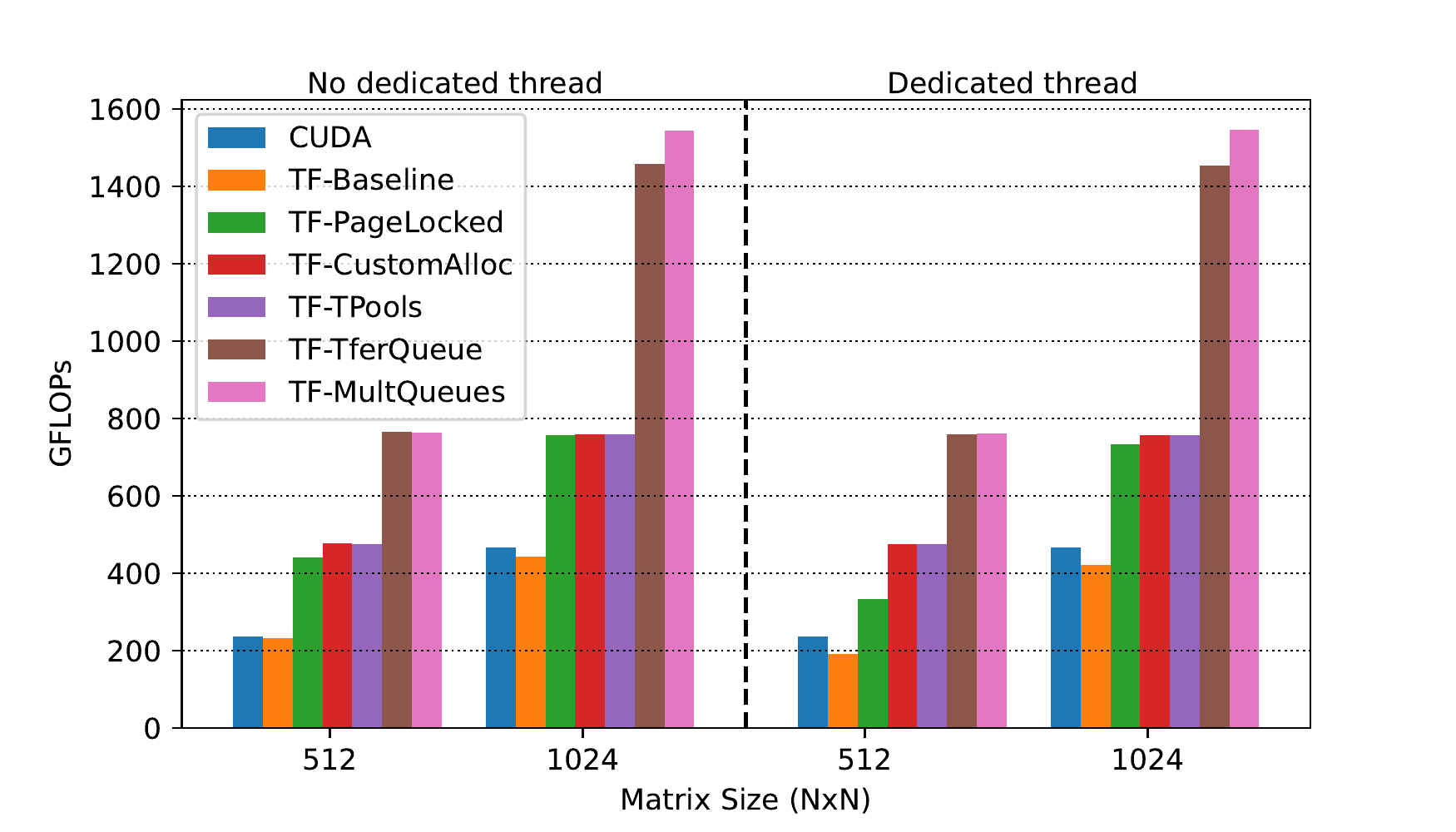}
    \caption{Performance of the heterogeneous tasking framework without (left) and with (right) a dedicated thread for the GPU device  on the matrix multiply benchmark for different matrix sizes}
    \label{fig:dgemm_benchmarks}
\end{figure}

So far, all the optimizations implemented were focused in improving the overlap of the CPU and GPU operations; however, operations that run in the GPU are still serialized by default. We enable implicit concurrency  between the different GPU operations by utilizing multiple execution streams provided by the device implementations (OpenCL command queues or CUDA streams for NVIDIA GPUs). Our implementation uses multiple streams for submitting computation kernels and two for memory transfer operations (one for each direction). 
The framework will guarantee that conflicting operations will never run in parallel based on user-provided dependencies and data access requirements. The results of this optimization are substantial in both implementations, attaining an improvement of 100\%, however, the implementation with the dedicated thread requires larger matrices to accomplish that (Fig.\ref{fig:dgemm_benchmarks}; TF-TferQueue, TF-MultQueues). 

\subsubsection{Other optimizations} 
To mitigate the effects of system calls and thread synchronizations, we have introduced request memory pools. Request pools are maintained per active thread and the memory of a request is recycled in the pool once the respective operations has completed. Another optimization regarding requests was implemented on the queues that are used to submit a request to a device. Initially, we used structures provided by the C++ STL, protected by a mutex to implement such queues. These queues were substituted with custom lists that avoid allocating nodes to store new elements. Moreover, the mutex locking step was moved after the queue's size was checked to eliminate unneeded locking operations. This optimization is less important than the ones discussed above, about 2\% improvement, but helps to attain a more consistent latency, especially when the dedicated thread is used (Fig.\ref{fig:dgemm_benchmarks}; TF-Tpools).  

The overall performance improvement achieved from this series of optimizations can exceed 300\% depending on the size of the matrices. Our framework offers all these optimizations with minimal involvement of the user and will continue to improve without any modifications required in the application code. Note that the performance improvements will get higher as the amount of memory transfers per computation decreases.

\subsection{Heterogeneous PREMA}
\begin{figure*}[t]
    \centering
    \includegraphics[width=0.32\linewidth]{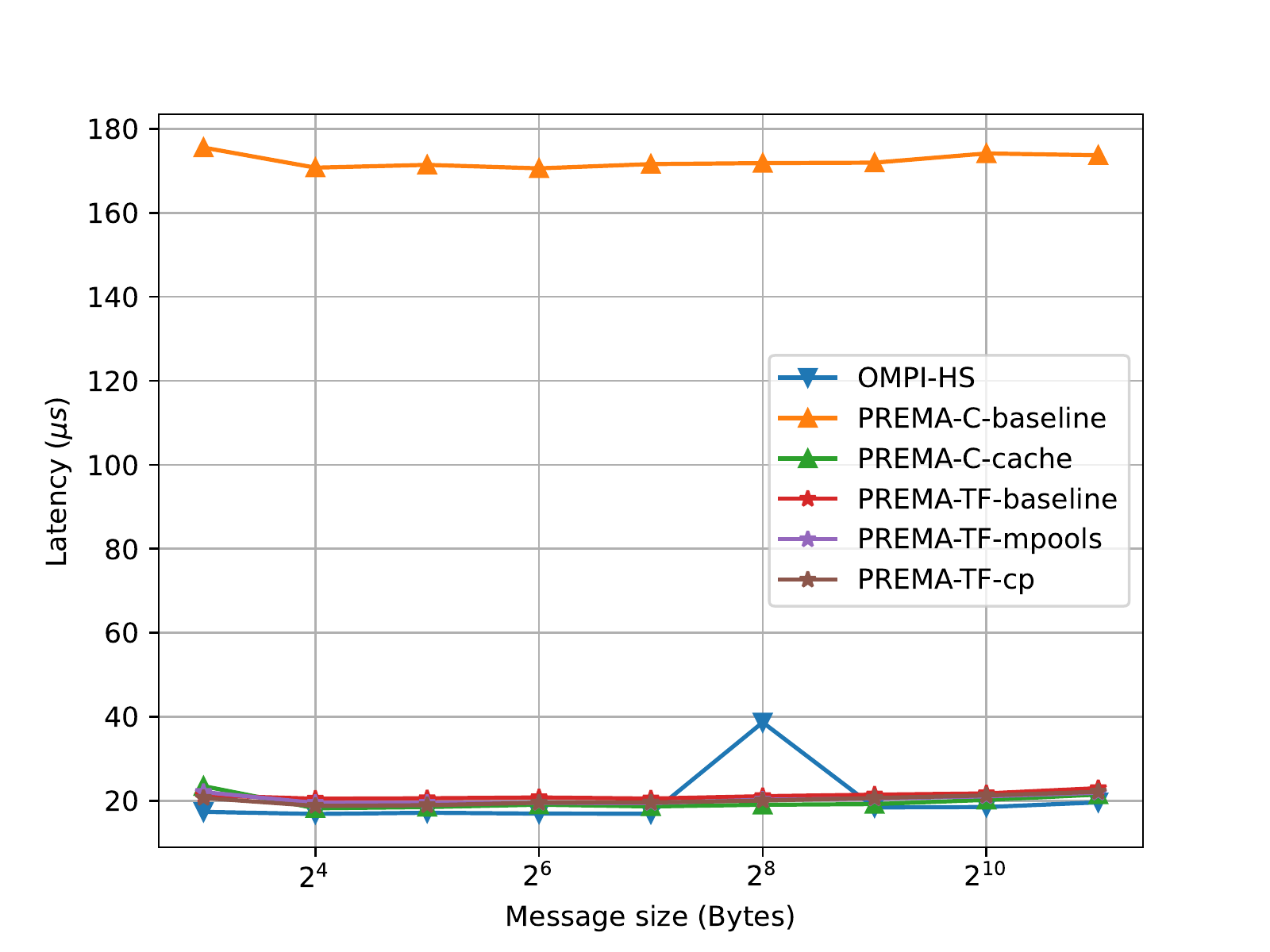}
    \includegraphics[width=0.32\linewidth]{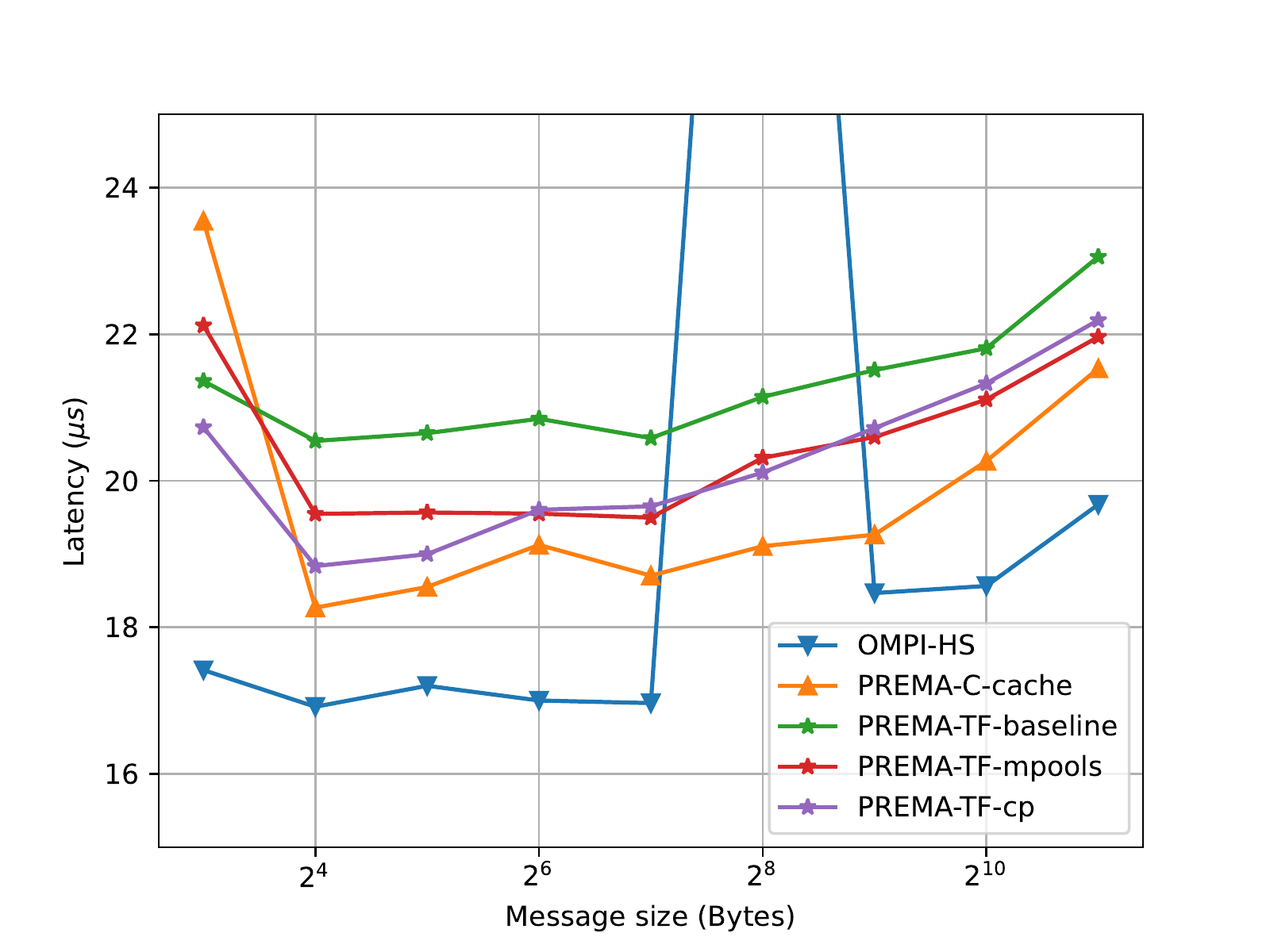}
    \includegraphics[width=0.32\linewidth]{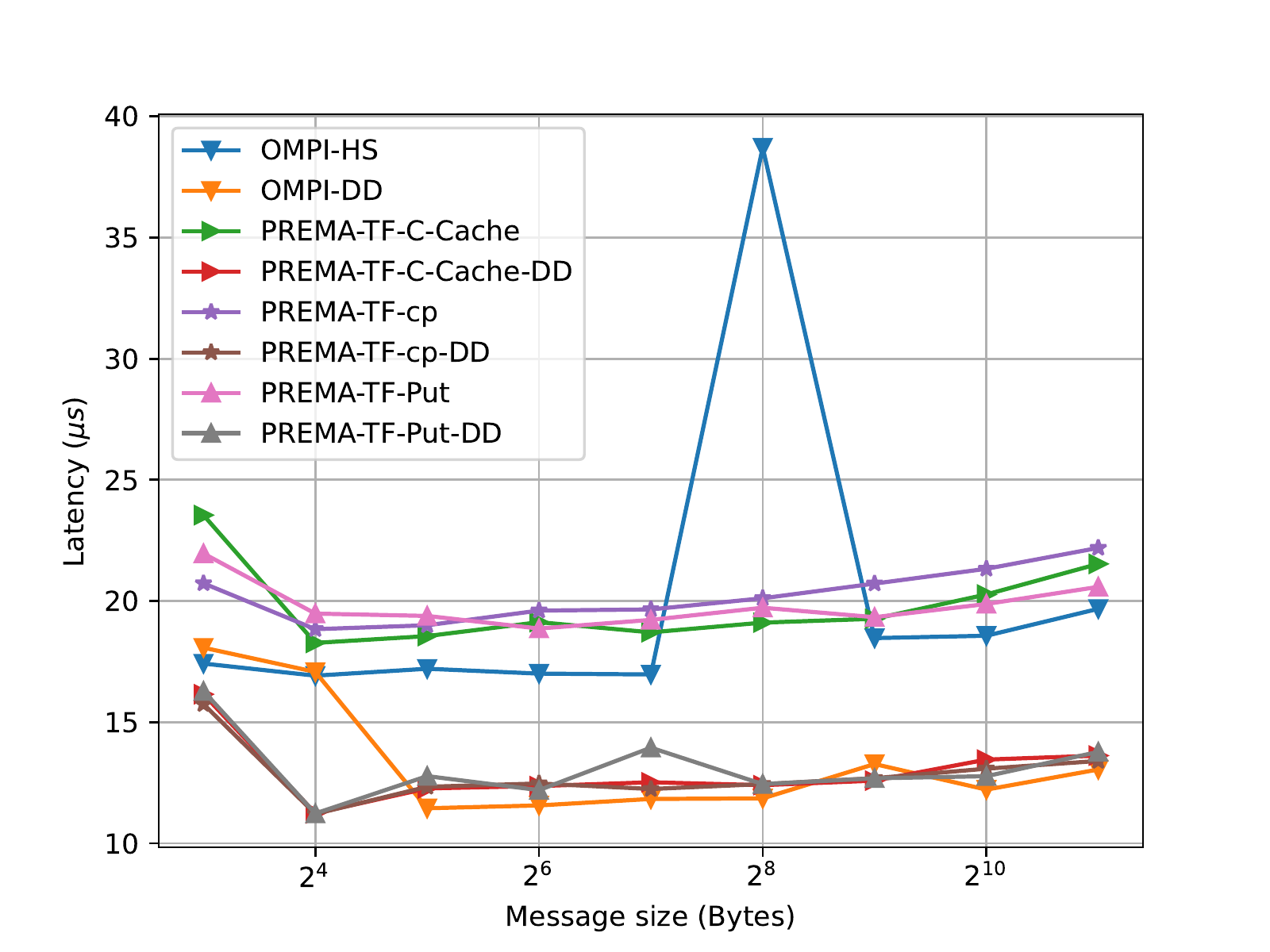}    
    \caption{Performance evaluation of different latency optimizations using a ping-pong benchmark ( \textbf{OMPI-HS/DD}: OpenMPI Implementation with host staging/direct-to-device, \textbf{PREMA-C}: PREMA using CUDA explicitly, \textbf{PREMA-TF}: PREMA using the tasking framework, \textbf{PREMA-TF-Put}: PREMA utilizing the tasking framework's put operation. A suffix of DD refers to implementations using heterogeneity-aware communication substrates.)}
    \label{fig:pp_opt}
\end{figure*}
\begin{figure*}
    \centering
    \includegraphics[width=0.35\linewidth]{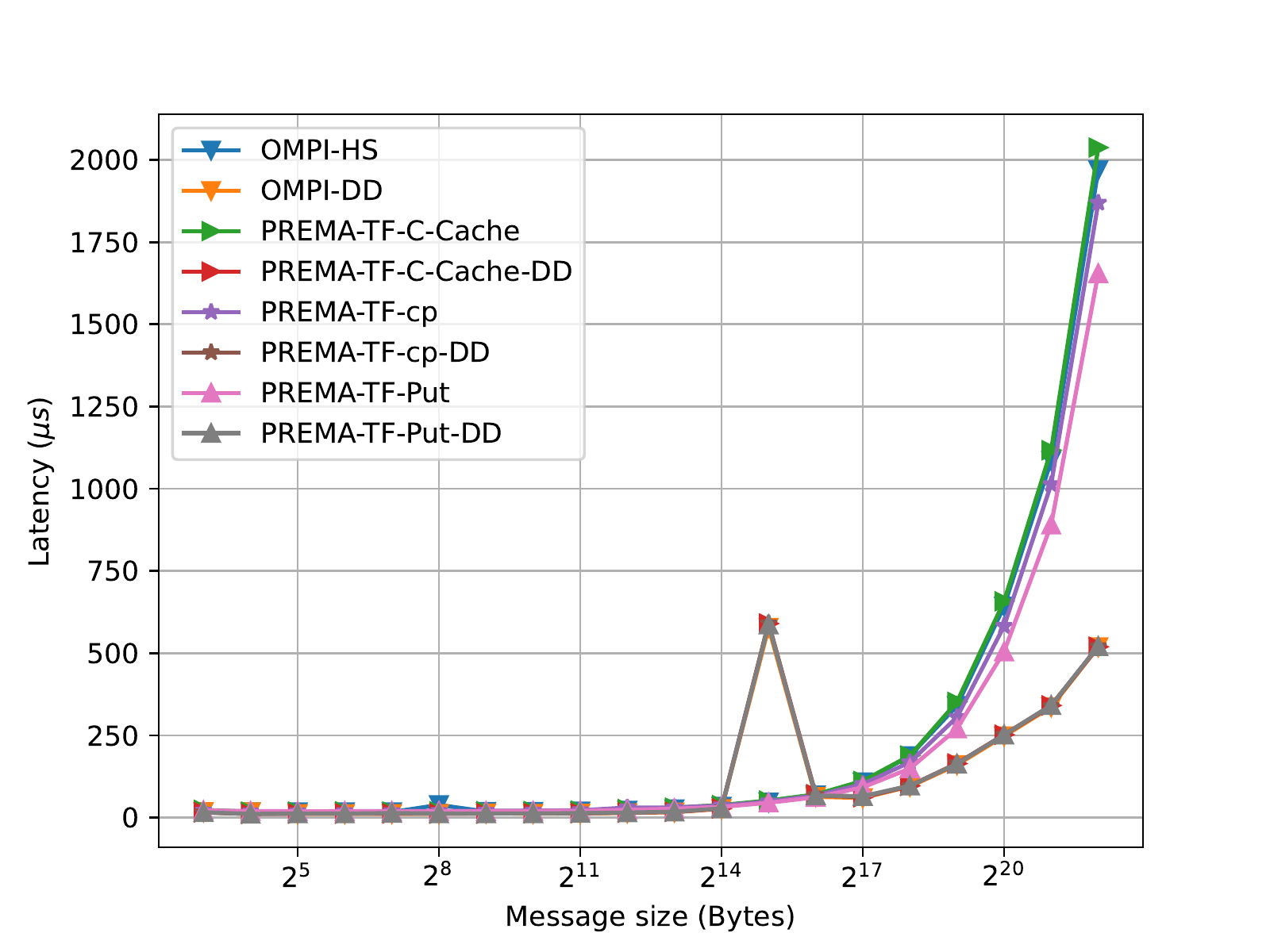}
    \includegraphics[width=0.35\linewidth]{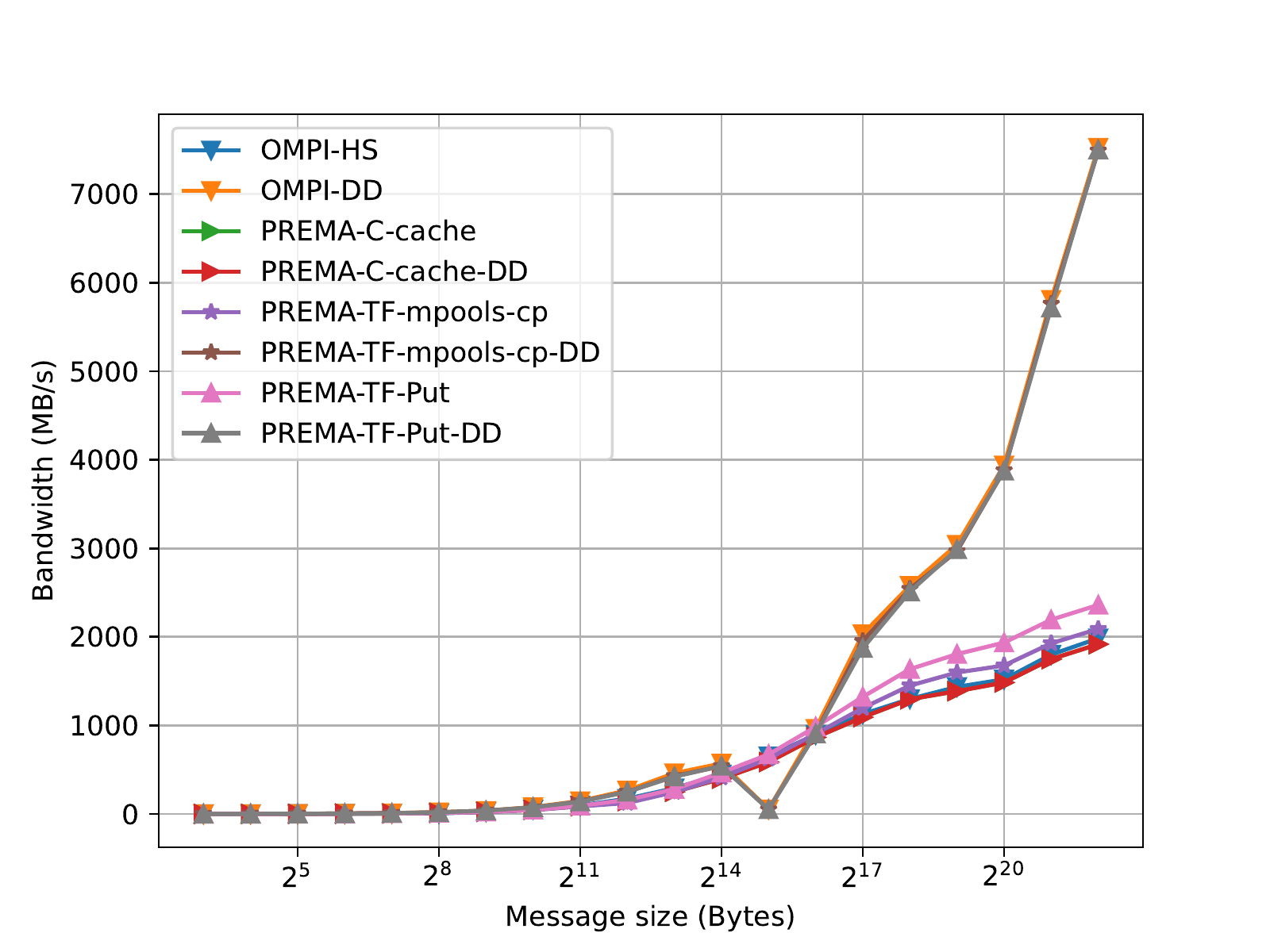}
    \caption{Overall latency and bandwidth of the optimized features provided by PREMA versus the MPI on a ping pong benchmark}
    \label{fig:pp_benchmarks}
\end{figure*}
To optimize the performance of the new heterogeneity-aware version of PREMA, we experiment with optimizations that can help us mitigate the overheads following the implementation of remote handler invocations that include heterogeneous memory both without and with the tasking framework (without a dedicated thread). We evaluate our optimizations on a simple ping-pong benchmark for inter-node communications and compare it with an OpenMPI (OMPI) implementation. Our benchmark runs a total of 100 ping-pong iterations with message sizes ranging between 8 bytes to 4 MBs and we report the average latency and bandwidth observed per message size.

\subsubsection{Device Message Receiving Cache}
In PREMA, messages are one-sided, asynchronous and are received implicitly to invoke a designated task on their target. The receiver cannot specify a memory region where the message buffer shall be stored (like MPI), so PREMA has to dynamically allocate memory to receive the incoming message buffer and provide it to the application handler invocation. In our initial implementation, without the tasking framework, simply allocating new device memory for each incoming message resulted in bad performance, increasing the latency experienced about up to ten times compared to the respective OpenMPI implementation (see Fig.~\ref{fig:pp_opt} left; PREMA-C-baseline). We were able to avoid this behavior but allocating a cache in the device specifically for the buffers of received messages. When a new message buffer is about to be received, memory is requested from the cache instead of the device API if possible. The cache allowed us to attain performance within 10\% of that achieved by the OpenMPI (Fig.~\ref{fig:pp_opt} left, middle; PREMA-C-cache). 

\subsubsection{Preallocating hetero\_objects}
Hetero\_objects automatically utilize memory pools for device memory, thus, implicitly overcoming the issue faced in the case where device memory is handled explicitly and achieving performance within 25\% of the OpenMPI. However, we can still improve some latencies caused by the constant allocation and deallocation of temporary hetero\_objects that wrap message buffers targeting device memory. Specifically, we found that the data structures allocated for a hetero\_object for bookkeeping different operations targeting the object in different devices can significantly affect communication performance. Since these structures can be allocated in advance, we use a pool of preallocated semi-initialized hetero\_objects to mitigate this effect. 
This optimization improved the latency experienced for small messages by about 5\%, bringing the performance of PREMA within 20\% of the OpenMPI, as can be seen in \ref{fig:pp_opt} (middle; PREMA-TF-mpools).

\subsubsection{Avoiding Message Buffer Copies}
The process of inter-node message transfers presented in \ref{sec:hetero_prema_impl} with host-staging includes an optimization for small buffer sizes to only send one message. Small messages with a size up to 512 bytes (header + buffer size) will include the actual application data/buffer appended in the end of the message. This helps optimize the performance of small messages where even negligible overheads are noticeable. However, when hetero\_objects are used this introduces an extra copy. As explained before, requesting access to the underlying data of a hetero\_object will implicitly copy the data to the host in a buffer maintained by the framework. In order to append the data at the end of the message header, PREMA needed to copy the data from the framework's location to the message header.

A new method is introduced in the hetero\_object API that allows the user to request a copy of its underlying data to a designated memory region at host. PREMA utilizes this feature to request from the framework to directly transfer the device data at the end on the message header buffer. This change provides another small improvement in the communication critical path and attains up to 5\% lower latency (Fig.~\ref{fig:pp_opt} middle; PREMA-cp). 

\subsubsection{Direct to Device Transfers}
The optimizations presented so far target the generic implementation of heterogeneity on top of PREMA where the communication library/hardware is not expected to be heterogeneity-aware. However, as already mentioned in detail in \ref{sec:hetero_prema_impl}, PREMA is able to leverage the capabilities of hardware/libraries that have been integrated with support for direct device-to-device communication. By following the procedure presented previously, the latencies observed for heterogeneity-aware hardware can be significantly mitigated. Figure \ref{fig:pp_opt} (right) shows the performance of the optimized version of each operation and the OpenMPI when using host-staging for communication, and when direct to device communication is possible. The attained performance in this case is up to 100\% better than the host-staging case for small messages (Fig.\ref{fig:pp_opt} right) and up to 200\% for large messages, as shown in Fig. \ref{fig:pp_benchmarks} left (latency) and right (bandwidth). 

Overall, all three operations introduced in PREMA to handle heterogeneity perform on par with the OpenMPI in terms of latency and bandwidth. It is interesting to notice that some of these operations, like the put operation, outperform OpenMPI for large messages (Fig.\ref{fig:pp_benchmarks} right; PREMA-TF-Put) by implicitly leveraging from the page-locked memory of hetero\_objects.

\subsection{Proxy Application: Jacobi3d}
\begin{figure}
    \centering
    \includegraphics[width=0.9\linewidth]{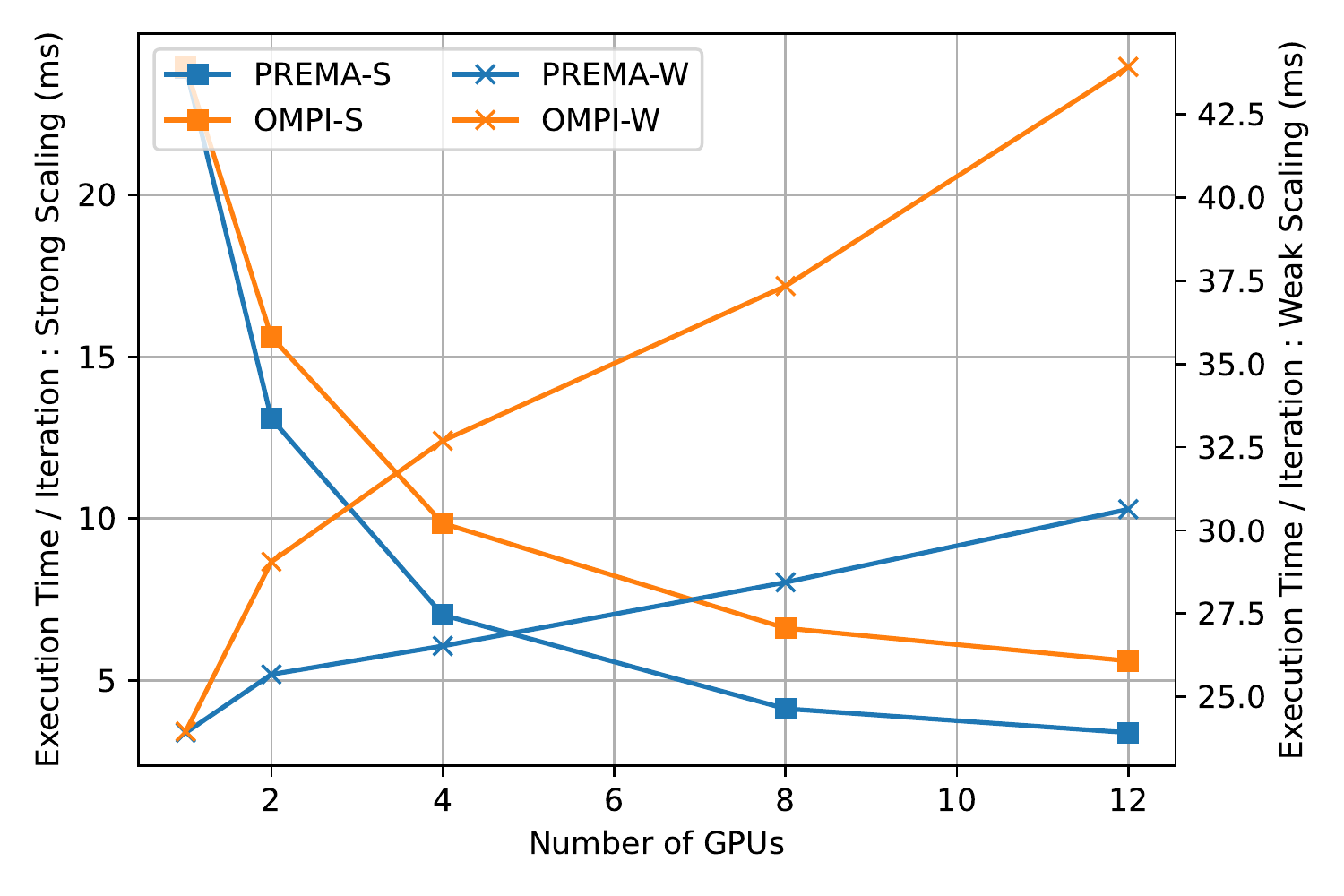}
    \caption{Strong (left y-axis) and weak (right y-axis) scaling performance of the Jacboid3D proxy application.}
    \label{fig:ev_jacobi3d}
\end{figure}
\label{sec:jacobi3d}
We have adapted the Charm++ proxy Jacobi3D application~\cite{jacobi3dCharm} in PREMA. The proxy performs a fixed number of iterations of the Jacobi method on GPUs in a 3D domain which is decomposed into cuboids and wrapped into mobile objects. In each iteration, the mobile objects exchange halo data which consists of packing the GPU data and transferring them to their respective neighbors. On the receiving side, the data are unpacked into the GPU and once all halos have been received the Jacobi update is executed on the GPU. 

Fig.\ref{fig:ev_jacobi3d} (left y-axis) shows the execution time of the heterogeneous PREMA versus the OpenMPI counterpart for a 1024x1024x768 domain (strong scaling). The implementation with PREMA achieves up to 40\% better performance. For weak scaling, the domain's size is increased according to the increase in number of distributed GPUs.  The performance achieved is up to 30\% (Fig.\ref{fig:ev_jacobi3d}; right y- axis) better than the OpenMPI implementation. The improvements observed stem from automatically overlapping message passing, host-device memory transfers and kernel invocations.

\section{Conclusion and Future Work}
\label{sec:conclusion}
This paper introduced an efficient tasking framework that handles performance portability for multi-device heterogeneous nodes. This framework automatically scales applications to multiple devices while handling efficient scheduling, load balancing, task dependencies, memory transfers and overlapping latencies, attaining a performance improvement of up to 300\%. Integrating this framework into a distributed runtime system resulted in a complete library that can leverage exascale HPC systems consisting of multiple heterogeneous nodes. This paper presents the design and implementation of this library, as well as evaluations and optimizations performed (up to 10 times improvement over a naive implementation). Evaluation results on a proxy application show that the end product of this work incurs low latency and scalable performance (up to 40\% versus the MPI+CUDA) while providing a simple and uniform interface, independent of the target hardware. In the future, we plan to further improve the performance of the tasking framework and provide compiler support to automatically generate device kernels. Moreover, we intend to extend PREMA's implicit load balancing layer to accommodate workload of heterogeneous devices and experiment with a range of load balancing and scheduling policies.  

\section*{Acknowledgment}
This work is funded in part by the Dominion Fellowship, the Richard T. Cheng Endowment at Old Dominion University and NSF MRI grant no: CNS-1828593. The authors would like to thank the ODU ITS group for their support in utilizing the University's HPC cluster.  

\bibliographystyle{ieeetr}
\bibliography{references}

\end{document}